# Microseconds, milliseconds and seconds: deconvoluting the dynamic behaviour of planar perovskite solar cells.

Adam Pockett[a]; Giles Eperon[b,c]; Nobuya Sakai[b]; Henry Snaith[b]; Laurence M Peter[a*] and Petra J Cameron[a*]

Perovskite solar cells (PSC) are shown to behave as coupled ionic-electronic conductors with strong evidence that the ionic environment moderates both the rate of electron-hole recombination and the band offsets in planar PSC. Numerous models have been presented to explain the behavior of perovskite solar cells, but to date no single model has emerged that can explain both the frequency and time dependent response of the devices. Here we present a straightforward coupled ionic-electronic model that can be used to explain the large amplitude transient behavior and the impedance response of PSC.

**Introduction**

The last five years have seen an unprecedented rise in the efficiency of metal halide perovskite solar cells. Certified efficiencies have now reached an impressive 22.1%,[1] and there has been a rapid expansion of research into halide perovskite materials (hereafter called 'perovskites') for photovoltaics, LEDs, lasers and superconductors.[2]

In the rush to make more efficient devices, understanding of the physical processes that underlie perovskite solar cells (PSC) is lagging behind. The consensus is that perovskites are ionic conductors with the possibility that both anion and cation vacancies are able to move through the material.[3-5] Activation energies of between 0.08 eV and 0.58 eV have been calculated for the migration of iodide vacancies.[6-9] The activation energy for migration of methyl ammonium vacancies is somewhat higher, with values between 0.46 eV and 0.84 eV reported in the literature.[6-9] A number of important cell parameters, in particular hysteresis in the current-voltage (jV) curves, have been attributed to mobile vacancies.[5] Other papers have suggested that it is the PSC contacts that are responsible for hysteresis. A number of different organic, inorganic and small molecule Electron Transporting Layers (ETL) and Hole Transporting Layers (HTL) were trialled in PSC and it was clear that the type of ETL and HTL strongly changed the degree of jV hysteresis and the overall device response.[10, 11] More recently there have been reports that both mobile ionic vacancies and contact dependent surface recombination are necessary to explain hysteresis in measured jV curves.[4, 5] The situation is further complicated by the fact that mobile ions may not be confined to the perovskite layer. In the case where polymer or small molecule contacts are used, leakage of ions through the organic contact has been reported.[12, 13] In the case of 'hard' inorganic contacts, mobile vacancies may become adsorbed/immobilised at the interface, introducing a surface dipole.[14] It also seems likely that mobile ionic vacancies play a part in the degradation of the perovskite material.[15]

The relationship between the behaviour of electrons and ions in PSC precludes the use of simple ionic or electronic models.[3] Research is increasingly showing how the instantaneous distribution of ionic vacancies following a stimulus (e.g. light, externally applied potential difference), can modify the magnitude of the current and voltage measured from a PSC.[5, 16] In fact the coupling of ionic and electronic effects could explain much of the anomalous behaviour of PSC including scan rate and temperature dependent hysteresis in the jV curves and a strong dependence of transient voltage and current measurements on any previous conditioning of the device. It is possible to imagine a number of ways that ionic vacancies could change the rate of charge injection and electron-hole recombination in PSC. Shielding of electronic charge by ionic vacancies could reduce recombination by decreasing the capture cross section for electrons and holes. The presence of specifically adsorbed ions at the perovskite-ETL and perovskite-HTL interfaces would introduce surface dipoles and modify barriers for electron/hole injection. The width of ionic double layers at the perovskite-ETL and perovskite-HTL interfaces could affect the rate of interfacial recombination as the charge carrier has to tunnel through the double layer.

Here we present a semi-quantitative model of PSC that explains how coupling between electronic and ionic charges is responsible for the dynamic behaviour of PSC. We first report photovoltage transients at a range of temperatures; photovoltage rise times are frequently used in the literature to diagnose PSC but differences in rise time and transient shape are poorly understood.[17] In the second part of this paper the response of PSC to small amplitude stimuli is reported. The coupling of ionic and electronic effects can be highly non-linear, making them difficult to probe using large amplitude measurements. In contrast small amplitude measurements, such as Electrochemical Impedance Spectroscopy (EIS) or Intensity Modulated Photovoltage Spectroscopy (IMVS) apply a very small perturbation to the system, effectively linearizing the response and making data interpretation much easier. We have measured EIS and IMVS at a range of temperatures and light intensities and find that the data is consistent with the perovskite acting as a coupled electronic-ionic conductor where the ionic environment modifies the rate of electron-hole recombination. We compare small amplitude measurements with the large amplitude photovoltage rise and Open Circuit Photovoltage Decay (OCVD) measurements. We demonstrate that a simple equivalent circuit model can explain all the results and show that activation energies for ion migration in PSC can be obtained from EIS, IMVS and OCVD. Our model also explains the observation that the jV response of a PSC is dependent on the nature of the hole or electron extracting contact as well as the presence of ionic vacancies in the device.

**Experimental**

A full description of the experimental techniques has been given previously.[18] Briefly, the study measured a number of planar perovskite solar cells (PSC; 9 pixels across 4 devices). The cells had stabilized efficiencies of η ≈11-12% and an active surface area of 0.119cm$^2$ (cells were masked to 0.092 cm$^2$ jV curves; see Figure S1 for typical response). The active layer was ~425 nm thick film of $MAPbI_{3-x}Cl_x$ sandwiched between spin coated $TiO_2$ and spiro-OMeTAD(2,2',7,7'-Tetrakis-(N,N-di-4-methoxyphenyl amino)-9,9'-spirobifluorene) layers, where the perovskite was deposited using a toluene-assisted drenching method as detailed previously.[19] The cells were characterised by EIS; IMVS and large amplitude photovoltage measurements (photovoltage rise/OCVD). All frequency and transient response measurements were performed using a Solartron Modulab XM electrochemical system with optical bench attachment. Impedance measurements were performed at different light intensities under open-circuit conditions over a frequency range of 1 MHz to 3 mHz. IMVS measurements were performed from 200 kHz to 3mHz utilizing the Modulab's reference photodiode to correct for phase lag and attenuation of the LED signal at high frequency. Illumination was provided by a blue LED (470 nm) at intensities between 73 mW cm$^{-2}$ and 1 mW cm$^{-2}$. The highest intensity produced a photocurrent equivalent to the 1 sun AM1.5 value. For the large amplitude transient measurements, the devices were left in the dark for several minutes until the residual $V_{oc}$ had dropped to less than 1 mV. The photovoltage rise was monitored until a steady state $V_{oc}$ had been achieved, before switching the light off and recording the decay. Samples were mounted in a desiccated cell holder that maintained a dry environment (<10% RH) and constant temperature using a Peltier element and controller. The temperature was varied between -25 $^o$C and + 35 $^o$C

**Results and Discussion**

**The Time Domain: Photovoltage Rise and Open Circuit Photovoltage Decay (OCVD).**

We characterised several planar perovskite cells using large amplitude photovoltage measurements. We record both the photocurrent rise time and the decay in the steady state photovoltage. The cell was allowed to equilibrate fully in the dark ($V_{oc}$ < 1mV) and then illuminated at open circuit. Photovolatge rise transients were recorded until a photostationary value of the open circuit voltage was achieved, and then OCVD was measured. The benefit of large amplitude transient measurements is that we can start from a well-defined initial condition (i.e. equilibrium in the dark for electrons, holes and ions). In OCVD, the decay in photovoltage can reflect a number of processes including discharge of the chemical capacitance (recombination of electrons and holes); discharge of the contact or junction capacitance or other slow depolarisation pathways such as the loss of alignment in ferroelectric domains.[20]

The physical processes that are being measured by OCVD in PSC are still being debated in the literature. In the case where there is simple pseudo first order electron-hole recombination, the rate of OCVD decay is related to the carrier lifetime, $\tau_n$.

$$\tau_n = -\frac{kT}{q}\left(\frac{dV}{dt}\right)^{-1} \quad (1)$$

Bertoluzzi *et al* suggested that OCVD for PSC can involve complex co-operative relaxation phenomena which could give rise to non-exponential relaxation.[20] The authors defined a more general OCVD 'relaxation lifetime', $\tau_{relax}$, where

$$\tau_{relax}(V) = \left(-\frac{1}{V}\frac{dV}{dt}\right)^{-1} \qquad (2)$$

In dye-sensitized solar cells (DSCs), Equation 1 gives the effective electron lifetime as electrons in the mesoporous $TiO_2$ recombine with acceptors in the liquid electrolyte or solid state hole conductor.[21, 22] In this case, $\tau_n$ is related to the discharge of the chemical capacitance $C_\mu$ across the recombination (back electron transfer) resistance, $R_{recomb}$. In DSCs the chemical capacitance is associated with trapped electronic charge in the mesoporous titania network. In our previous studies, we showed that the microsecond to second time scale OCVD responses for PSC do not contain information on free carrier lifetimes.[18] This is unsurprising since photoluminescence measurements show that electron-hole recombination takes place on the nanosecond timescale.[23, 24]

In common with Baumann et al. we also observed a 'persistent' or long lived photovoltage with a change in slope at t ≈ $10^{-2}$ s.[25] Here we have measured temperature dependent photovolatge rise and OCVD for the first time. Photovoltage transients were also measured as a function of light intensity and as a function of illumination time. The highest light intensity was chosen as it gave photocurrents similar to those obtained at 1 Sun and AM 1.5. Figure 1(a) shows photovoltage transients for a planar device and Figure 1(b) is a close-up of the voltage rise measured for a PSC at a controlled temperature of 25°C. ln[1/(V-$V_{OC}$)] is plotted against time allowing easier comparison of the rise with the OCVD. Figures 1(c) and (d) show OCVD decay curves for the same PSC. One important point is that it is only possible to get perfect reproducibility in repeat photovoltage transient measurements if the cell is allowed to equilibrate fully in the dark before the light is switched on (< 1mV remaining voltage). The illumination time is also critical to obtaining reproducible photovoltage decay curves as again the cell has to reach the photostationary state before the light is switched off again. For some of the cells studied here, this required illumination for several minutes. The high reproducibility of our voltage rise and decay measurements is illustrated in Figure S2.

It is immediately clear from Figures 1(b) and 1(c) that both photovoltage rise and the $V_{OC}$ decay have a fast and a slow component with associated lifetimes that we will designate $\tau_{rise,1}$ (fast); $\tau_{rise,2}$ (slow); $\tau_{OCVD,1}$ (fast) and $\tau_{OCVD,2}$ (slow). When the light was switched on at the highest intensity, the $V_{OC}$ rose rapidly to ~0.55V; before increasing slowly towards a maximum steady state photovoltage of nearly 1V over the next 2-3 minutes. When the light was switched off there was a fast drop in $V_{oc}$ to ~ 0.5V followed by a slow decay. As is clear from Figure 1(d) the fast initial $V_{oc}$ decay to ~0.5V occurs between $10^{-6}$ s and $10^{-3}$ s at 25°C. In comparison the slow components of both the rise and decay are on the second time scale. In between the fast and slow decay, a small recovery of the $V_{OC}$ was observed on the ms timescale. Figure 1(f) shows the intensity dependent 'bounce back' in the photovoltage. The 'bounce back' was most visible at the highest light intensity used (e.g. 73 mWcm$^{-2}$) and is therefore relevant to cells operating close to 1 Sun. At this intensity the voltage *actually increased* by ~ 10 mV before it started to decay again.

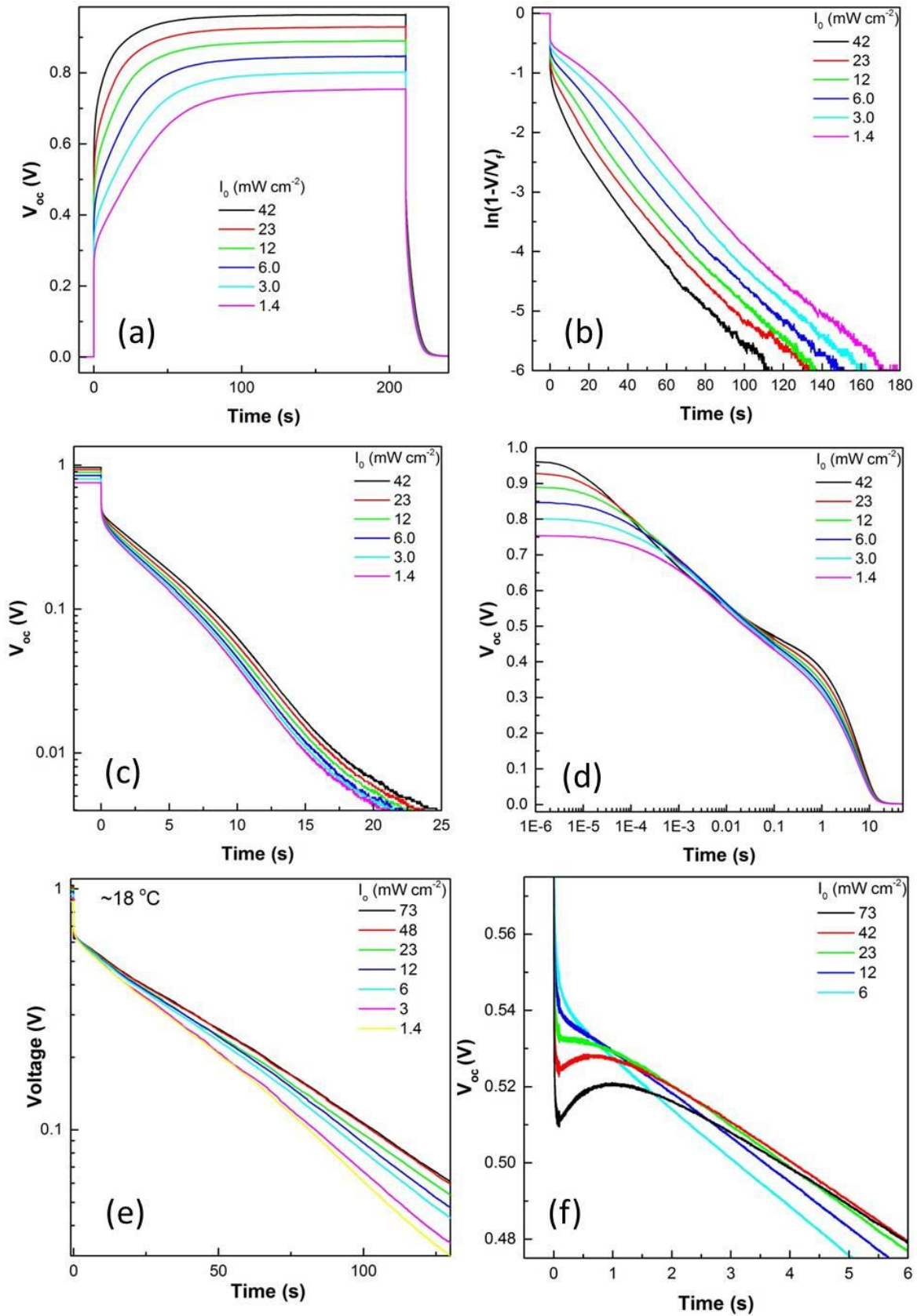

Figure 1 (a) Large amplitude photovoltage transient for cell CT-3 at 25°C. (b) Semi-logarithmic plot of the photovoltage rise as a function of light intensity (c) OCVD decay plotted on a semi-logarithmic plot (d) OCVD decay from (c) plotted versus log time. (e) OCVD curves showing the extraordinary linearity in the $V_{oc}$ decay at long times for a different cell measured at 18°C. (f) a close up of the transition between the fast (µs) decay and the slow (s-min) parts of the OCVD curves showing an intensity dependent 'bounce-back' of the voltage which was observed for many of the cells at the transition between the fast and slow decay.

The room temperature photovoltage transients reveal the considerable complexity of the relaxation processes occurring in the PSC. At the start of the experiment the cell is fully relaxed in the dark and ionic double layers are present in the perovskite at the contacts with the ETM and HTM (Figure 2). It has been shown previously that the defect density in perovskite films is sufficient to create compact double layers at the interfaces[5]. When the contacts are added to a PSC in the dark, anion or cation vacancies can move to create ionic double layers at the perovskite/electron extracting and perovskite/hole extracting interfaces. A simple calculation (see ESI) suggests that a concentration of mobile defects of the order of $10^{18}$ -$10^{19}$ cm$^{-3}$ would be sufficient to fully compensate a built in voltage of 1 eV across a 500 nm film and would mean that there was little or no electric field in the bulk perovskite. When the light is switched on, electronic charge appears almost immediately in the contacts (on the timescale of the electron-hole lifetimes, i.e. ns) and a photovoltage of ~0.5V is measured. The appearance of the photovoltage creates an electrical field in the film that is opposite in sign to the original built in voltage and will cause charged vacancies in the double layers to move towards the bulk, effectively discharging the double layer capacitance until a new equilibrium for ionic species is established. If the new ionic distribution is favourable – i.e. if the ions move in such a way to reduce the rate of recombination, then the $V_{oc}$ will continue to rise until the ionic atmosphere has fully relaxed. A decrease in recombination can be explained by shielding of electrons and holes by vacancies of opposite sign, reducing the capture cross section. However, a ~0.5V increase in the $V_{oc}$ would require bulk or surface recombination to be reduced by many orders of magnitude. In this situation it seems unlikely that shielding alone can explain the reduction in recombination rate, so it is necessary to also consider how changes in local band bending would modify the electron and hole concentration profiles (see discussion in the ESI and Figures S3-S5). Links between the ionic environment and recombination will also be discussed in the next section.

The OCVD decay measurement showed a fast initial decay on the μs-ms timescale that is consistent with discharge of the geometric capacitance across the recombination resistance; e.g. $\tau_{OCVD,1}$ (fast) = $R_{recomb}C_{geo}$. As we have shown previously[18], the fast decay has a time constant that closely matches that for $R_{recomb}C_{geo}$ measured using IMVS/EIS. This is confirmed for the cells measured here in Figure S6, which shows the remarkable similarity between $\tau_{OCVD,1}$ (calculated using equation 1) with $\tau_{HF}$ from EIS (see below).

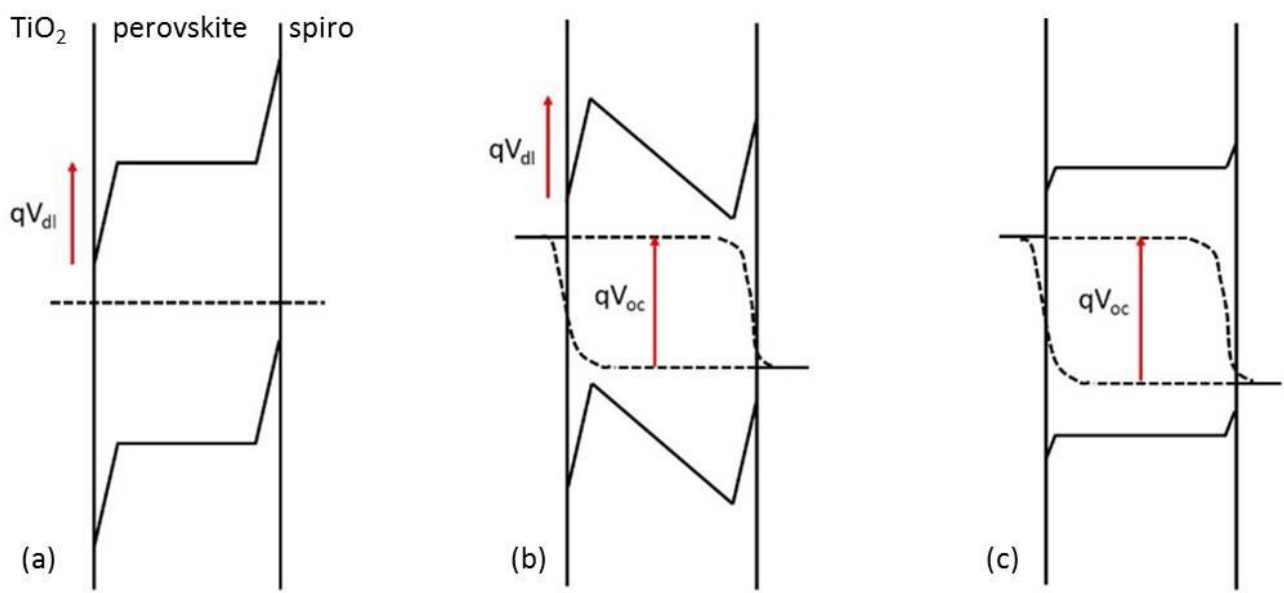

Figure 2. Sketch suggesting the changes in band bending induced by illumination of a cell where ion vacancies as wells as electrons and holes are in equilibrium. The photo-induced field leads to discharge of the double layers as ions return to the bulk form the near surface regions. (a) after equilibration of vacancies in the dark (b) situation immediately following illumination at open circuit (c) after equilibration of ionic vacancies at open circuit under illumination.

The slow component of the OCVD decay has a $\tau_{OCVD,2} \approx 5s$ for the same cell at 73 mW cm$^{-2}$ and 25 °C. The OCVD lifetime was extracted assuming a single life time and mono-exponential decay. Figure 1(c) shows that the OCVD is close to being linear on a semi-logarithmic plot. Furthermore all the OCVD plots became increasingly linear at lower temperatures (see below) making us believe that fitting a single exponential decay is justified. Additionally many of the cells (e.g. see Figure 1(e)) showed a highly linear response even close to room temperature.

In a simple PV model (Fig S7) the voltage decays as the geometric capacitance is discharged across the recombination resistance and the shunt resistance. Previously we have measured shunt resistances of the order of $10^5$ Ω for PSC.[18] Taking $R_{recomb}$ = 10-50 Ω (from EIS below) it is possible to model the OCVD curve that would be expected if all recombination was via $R_{recomb}$ and $R_{shunt}$ (Figures S8 and S9). It can be seen that for a shunt resistance of $10^6$ Ω or below the OCVD would be expected to be completely over in less than 1 s (below 100 ms for $10^5$ Ω). The observed response takes tens of seconds and unphysically high values of the shunt resistance need to be used to model such a slow decay. It is more likely that the slow part of the OCVD is measuring charging of the ionic double layer capacitance. This would suggest the equivalent circuit model for PSC shown in Figure 3.

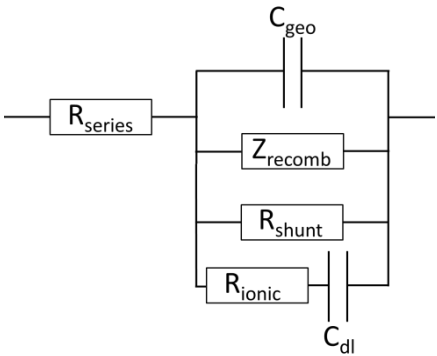

Figure 3 Full equivalent circuit model for PSC. $R_{shunt}$ is the shunt resistance, $R_{ionic}$ is the ionic resistance to ion movement and $C_{dl}$ is the ionic double layer capacitance at the perovskite-ETL and perovskite-HTL contacts, and $Z_{recomb}$ is the frequency dependent recombination resistance.

If $\tau_{OCVD,2} = R_{ionic}C_{dl}$, then taking a time constant of 5 s (at 25°C, 73 mWcm$^{-2}$) and a calculated double layer capacitance of 40 μF cm$^{-2}$ (see ESI) would give an ionic resistance of 1.25 x $10^5$ Ω cm$^2$. Eames et al have calculated mobilities of 4 x $10^{-11}$ cm$^2$ V$^{-1}$ s$^{-1}$ for iodide vacancies. If the defect density is calculated from this mobility and the calculated ionic resistance (1.25 x $10^5$ Ω cm$^2$), it would predict a very reasonable density of $10^{19}$ cm$^{-3}$ iodide vacancies.

To investigate further, large amplitude photovoltage transients were measured at a range of temperatures and the results are shown in Figure 4. It can be seen that the rise and fall in $V_{oc}$ is extremely slow at low temperatures, at -25°C the rise to a steady photovoltage takes 1800 s. When the light is switched off, the $V_{oc}$ has only decayed back to ~0.3V after 600s. $\tau_{rise, 2}$ and $\tau_{OCVD, 2}$ were extracted by fitting a single exponential and plotted as a function of temperature. A value of $E_A$ = 0.42 eV was extracted from the rise time and of $E_A$ = 0.58 eV from the decay time. Our experimental activation energies are remarkably close to the value calculated by Eames et al (0.58 eV)[6] for iodide vacancies. We therefore conclude that the slow component of OCVD curves is indeed measuring charging of the ionic double layer.

As discussed above, many cells showed a 'bounce back' in the $V_{OC}$ decay curves which occurred at the transition from the fast initial decay to the slow decay. The feature was highly dependent on light intensity and could not always be clearly resolved; but was visible in the majority of the cells measured for this study. At the highest intensity the voltage actually increased by ~ 10 mV before it started to decay again.

The mid-range 'bounce back' is also temperature dependent and became broader as the temperature decreased (Figure 4(c)). A recovery of the $V_{OC}$ strongly suggests that ionic vacancies can move to change the recombination resistance on the ms time scale. To try and better understand the μs, ms and s responses of PSC, small amplitude measurements were carried out using electrochemical impedance spectroscopy and intensity modulated photovoltage spectroscopy. The results will be discussed in the context of the large amplitude results below.

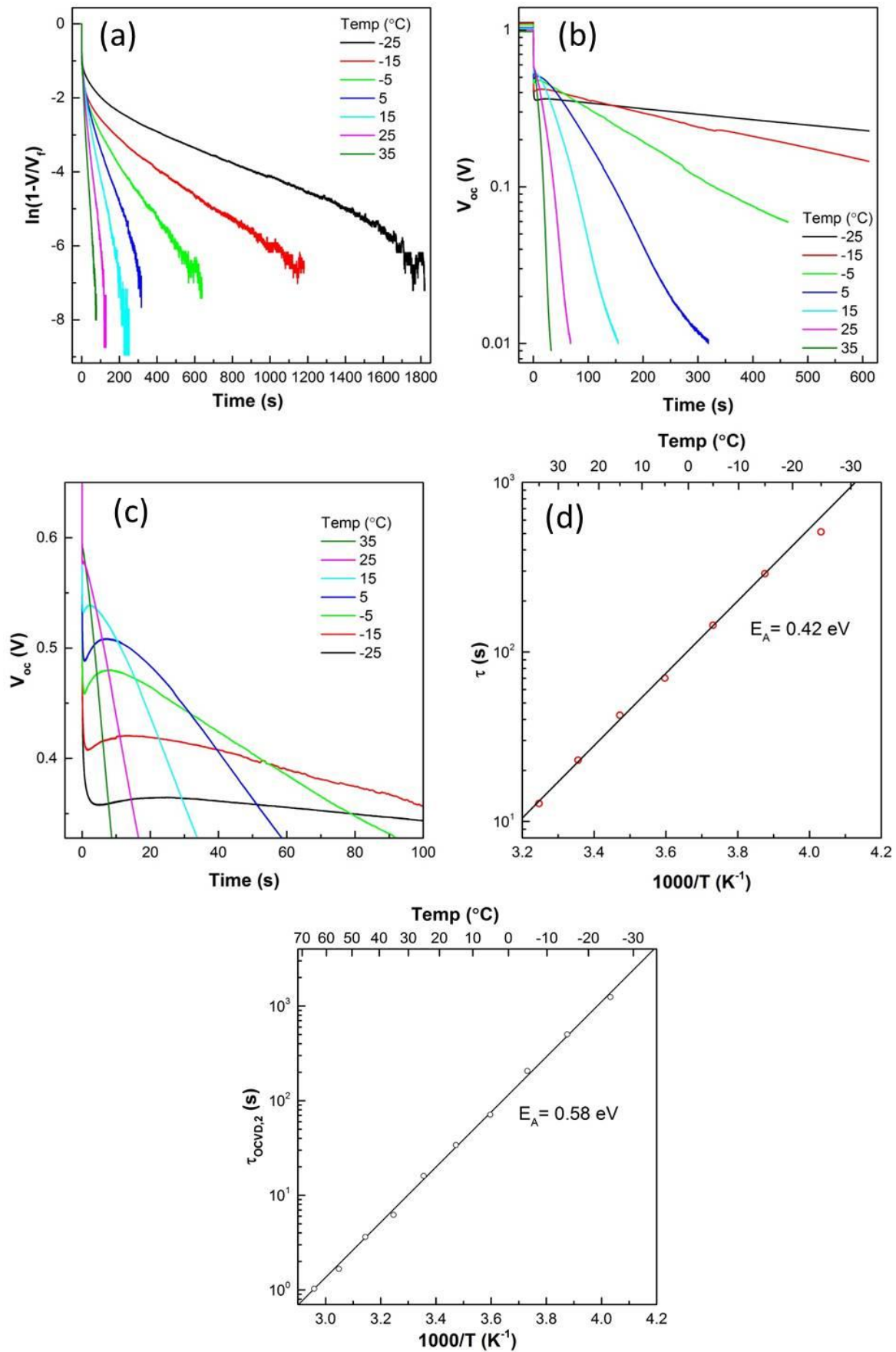

Figure 4. (a) Temperature dependence of the photovoltage rise. (b) temperature dependence of the photovoltage decay (c) the temperature dependence of the 'bounce back' (d) $\tau_{rise,2}$ as a function of temperature giving an activation energy connected with the slow rise in photovoltage. (e) $\tau_{OCVD,2}$ as a function of temperature giving the activation energy connected with the slow part of the OCVD curve.

**Response in the frequency domain: EIS and IMVS**

Over the last few years Electrochemical Impedance Spectroscopy (EIS) has emerged as a popular technique to study PSC. It has been used to characterise planar and mesoscopic PSC devices with a range of organic and inorganic electron and hole extracting contacts.[10, 11, 14, 26-35] Intensity modulated photocurrent spectroscopy (IMPS) and intensity modulated photovoltage spectroscopy (IMVS) are frequently used as complementary techniques to study Dye Sensitized Solar cells (DSC).[36-38] These techniques have been used much less often to study PSC;[18, 20, 30] largely because PSC behave very differently to DSC and the models for DSC cannot be applied.[18] In general, reported EIS spectra for PSC show a single high frequency semi-circular response in the dark with an extra low frequency semi-circle appearing under illumination. The form of the EIS response is remarkably similar for PSC with organic contacts, inorganic contacts as well as 'hysteresis free' devices. The high frequency response has been modelled by a resistance and capacitance in parallel – the resistance has been attributed to an electron transport resistance, an ion transport resistance or a recombination resistance. The capacitance has been linked to the geometric (also called the contact/junction) capacitance of the cell; the chemical capacitance, or a capacitance due to dipole depolarisation in the perovskite material.[11, 18, 31, 39-41] Interpretation of the low frequency response is even more widely debated. It has been attributed to the impedance of trap states,[35] an impedance due to degradation in the device, a 'giant dielectric' effect,[27, 28, 42] an impedance due to electron accumulation at the contacts[11, 43] and ionic diffusion.[32] Some PSC have shown a third feature at intermediate frequencies, e.g. Guerrero et al have reported an inductive loop in cells with efficient $SnO_2$ contacts.[11]

Here we have measured EIS spectra between 1 MHz and 3 mHz at open circuit under illumination. Figure 5(a) shows the EIS response at the highest light intensity used (73 mW cm$^{-2}$). Two semi-circles are clearly visible in Figure 1(a) with time constants that will be denoted $\tau_{HF}$ (high frequency (HF) feature) and $\tau_{LF}$ (low frequency (LF) feature). At 73 mWcm$^{-3}$, $\tau_{HF}$ is in the μs region. In our previous work,[18] we did not report measurements below 1 Hz as we were concerned about the stability of the cells when they were being measured at open circuit for long periods of time. In the present study, the cells were held in a dry and temperature controlled environment and the $V_{oc}$ values were extremely stable over the entire set of measurements. The low frequency time constant measured by EIS, $\tau_{LF}$, occurred on the second time scale. Interpretation of the high and low frequency responses will be discussed in the context of the IMVS results below.

Figure 5(b) shows the IMVS response for the same cell in the range 0.2 MHz to 3 mHz. In IMVS changes in open circuit voltage are measured as the PSC is illuminated with a sinusoidally varying light source. Data is shown at three light intensities: 73, 42 and 23 mW cm$^{-2}$. In common with EIS, high and low frequency semi-circles were observed with time constants designated $\tau_{HF}$ and $\tau_{LF}$. In addition a new mid-frequency semi-circle was also clearly visible with a time constant, designated $\tau_{MF}$ (mid frequency (MF) feature), on the ms time scale.

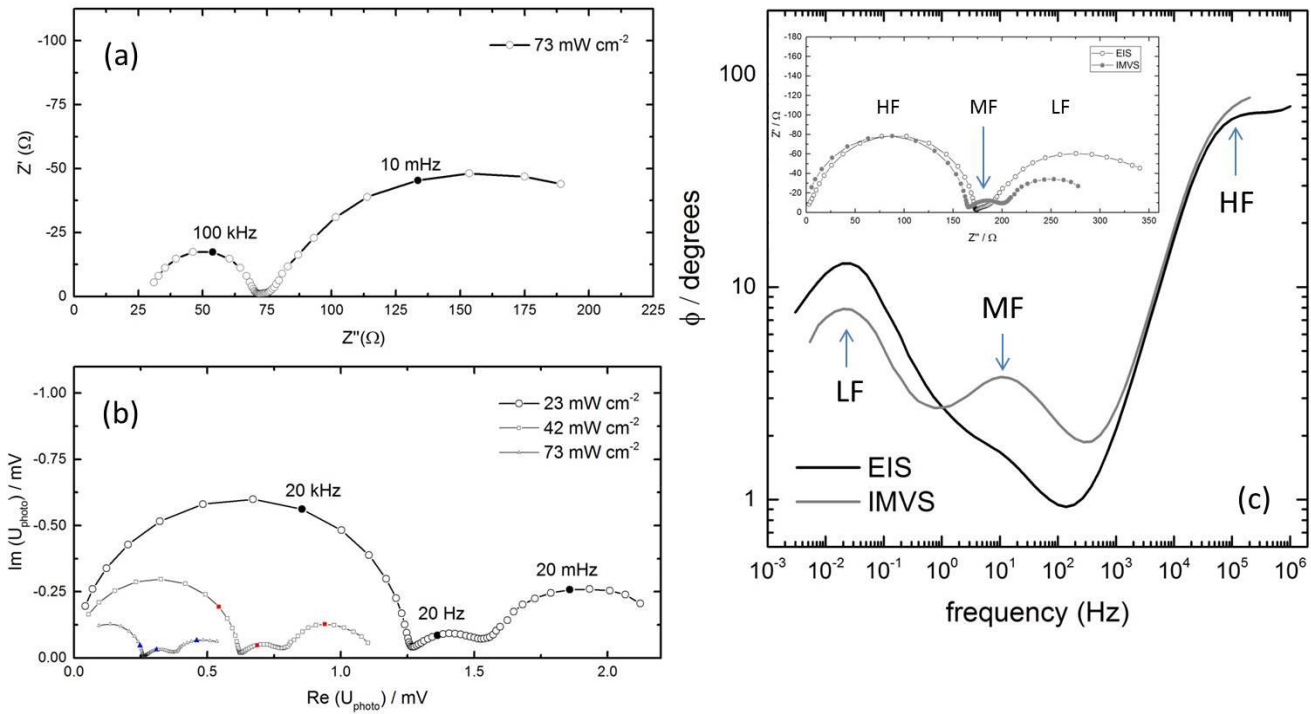

Figure 5. (a) Nyquist plot showing EIS at 73 mW cm-2; (b) IMVS at three different light intensities; (c) Bode plot showing a comparison of EIS and IMVS, where the IMVS has been converted by dividing by the ac photocurrent at Voc obtained using an IMPS measurement. Inset shows the corresponding Nyquist plot. All data was obtained for cell CT-4. The features have been labelled HF (high frequency); MF (mid-frequency) and LF (low frequency) to help layer discussion in the main text.

EIS and IMVS are related techniques. In EIS the Fermi-level is modulated externally by a sinusoidally varying voltage, in IMVS the Fermi-level is modulated internally by a sinusoidally varying light source. The main difference is that in EIS the current is measured in the external circuit. In IMVS no current is extracted and the $V_{OC}$ is measured. In principal the two techniques should probe the same processes occurring within the PSC. The $\tau_{HF}$ values measured from EIS and IMVS are very similar and show an identical dependence on light intensity (see Figure S11). As a result, we conclude that both techniques measure the same HF process. Previously we showed that the high frequency semi-circle measured by EIS was related to the geometric (also called junction or contact) capacitance of the PSC and the recombination resistance ($\tau_{HF}=C_{geo}R_{recomb}$).[18] Our evidence for this conclusion was linked to the dependence of the capacitance on light intensity and photovoltage. The capacitance was practically constant across the range of voltages measured which is the expected response of the geometric capacitance (with values 130-200 nF cm$^{-2}$; see also results from this study in Figure S12). A chemical capacitance, due to the accumulation of free carriers in the perovskite, was not observed. The chemical capacitance would have been expected to increase exponentially with voltage (see ESI of reference 17). In addition the value of the capacitance was entirely consistent with the thickness and dielectric constant of the perovskite layer in the PSC (C = $\varepsilon\varepsilon_0 A/d$). Our interpretation was subsequently confirmed by Guerrero et al who measured EIS on PSC containing perovskite layers of different thickness and showed that the high frequency capacitance scaled linearly with the inverse of the perovskite film thickness and was therefore due to the geometric capacitance.[11] We attributed the accompanying HF resistance to $R_{recomb}$ as it scaled linearly with light intensity (see ESI of Ref 18 and Figure S11). Guerrero et al also showed that the recombination resistance dominated when the perovskite film was thinner than 500 nm. We therefore conclude that in typical PSC devices the resistance extracted from the HF semi-circle measured by EIS is indeed a measure of recombination.

To underline the similarities between EIS and IMVS, it is instructive to relate the data directly. A direct comparison can be carried out in two ways. Firstly, IMVS data can be converted to EIS data by dividing the IMVS response by the current response to the same ac light stimulus. The ac voltage is divided by the ac current obtained from the low frequency intercept of an intensity modulated photocurrent (IMPS) measurement made at $V_{oc}$. The low frequency intercept gives the ac current in phase with the incident illumination. It is important to note that using the values from the more usual IMPS measurement at short circuit gives incorrect results as the photocurrent is overestimated. The second method of scaling the IMVS is to assume that the measured geometric capacitance of the cell is the same at the same light intensity in both measurements. This is reasonable as the high frequency response of both IMVS and EIS gives $\tau_{HF}=C_{geo}R_{recomb}$ with near identical values of $\tau_{HF}$. A simple scaling technique is therefore to scale the IMVS curve by a multiplier that gives the same value of $C_{geo}$ as obtained from the EIS measurement. Both scaling methods gave very similar results and both methods have been used in this study.

A direct comparison of EIS and IMVS spectra at 23 mW cm$^{-2}$ is given in the Nyquist and Bode plots in Figure 5(c). Since the series resistance is not measured by IMVS (no current is extracted through the external circuit in IMVS), it has been subtracted from the EIS response. The spectra look very similar. The low frequency time constants, $\tau_{LF}$, from EIS and IMVS are both in the second regime. At 72 mWcm$^{-2}$, $\tau_{LF}$ from EIS is ~28 s and from IMVS is ~12s. Figure 5(c) further shows that there is a 'hidden' mid-frequency time constant in the EIS measurement which is only just visible on the Bode plot when it is (unusually) plotted as log($\phi$/degrees) versus frequency. It is interesting that the IMVS response is more sensitive to the mid-frequency feature. As mentioned above, in IMVS no current is extracted in the external circuit as the voltage is measured as a function of the modulated illumination intensity (the Fermi-level is modulated internally by light). In EIS, current is measured as a function of the modulated voltage (Fermi-level is modulated externally by voltage). It appears that extraction of charge in the EIS measurements may lead to a 'smearing out' of the mid frequency response. Nevertheless, the results clearly show that both EIS and IMVS are measuring the same three relaxation processes that occur on the μs, ms and s timescales.

To further investigate the HF, MF and LF features, values of the resistance were extracted using a simple semi-circle fit. The values were then plotted as a function of light intensity. Figure 6(a) shows the intensity dependence of the resistance values obtained from the HF and LF semi-circles in the EIS measurements. The recombination resistance extracted from the HF semi-circle shows a linear dependence on intensity. We have shown previously that a linear dependence of $R_{recomb}$ on light intensity is expected for PSC (see ESI of Ref 18)[18] Significantly, the LF resistance shows *an almost identical dependence* on light intensity (Figure 6(a); gradient 0.94 Ω mW$^{-1}$cm$^2$ (LF) and 0.97 Ω mW$^{-1}$cm$^2$ (HF); $R^2$ > 99% in both cases). Figure 6(b) shows the intensity dependence of the resistances extracted from the HF and MF semi-circles observed in the IMVS data for a different cell. The data show that the MF time constant also shows a linear dependence on light intensity, albeit with a small deviation from linearity at the highest intensity where the data was hardest to fit accurately. In general it was harder to obtain accurate values for the MF semi-circle as it merged with the LF and HF response. Despite this, the intensity dependence of the HF and MF features remain similar (gradients -0.85 Ω mW$^{-1}$cm$^2$ (HF) and -0.76 Ω mW$^{-1}$ cm$^2$ (MF); $R^2$ > 99%). At lower intensities (where the fitting was easiest, the gradient of the MF plot increases to -0.8 Ω mW$^{-1}$cm$^2$.

We have also measured the temperature dependence of $\tau_{MF}$ and $\tau_{LF}$ by measuring EIS and IMVS spectra of a cell held in an inert atmosphere at temperatures between -25 and 35 °C. Time constants measured from spectra taken at different temperatures are shown in Figure 6(c) and (d). It is clear that both MF and LF time constants are temperature dependent. The Arrhenius plots give activation energies of 0.55 eV and 0.68 eV for the MF and LF processes respectively. Values of the corresponding 'attempt frequencies' obtained from the pre-exponential factors (neglecting entropic effects) were 1.8 x 10$^{11}$ s$^{-1}$ and 1.2 x 10$^{10}$ s$^{-1}$ respectively. An activation enthalpy of 0.55eV for the movement of iodide vacancies is also close to the value calculated by Eames *et al* (0.58 eV)[6]. The attempt frequency of 10$^{11}$ Hz is also reasonable. Eames *et al* suggest a typical attempt frequency of 10$^{12}$ Hz for ionic species at a temperature of 320 K – our measurements were carried out at the slightly lower temperature of 298 K. This strongly suggests that the MF feature is connected with movement of iodide vacancies.

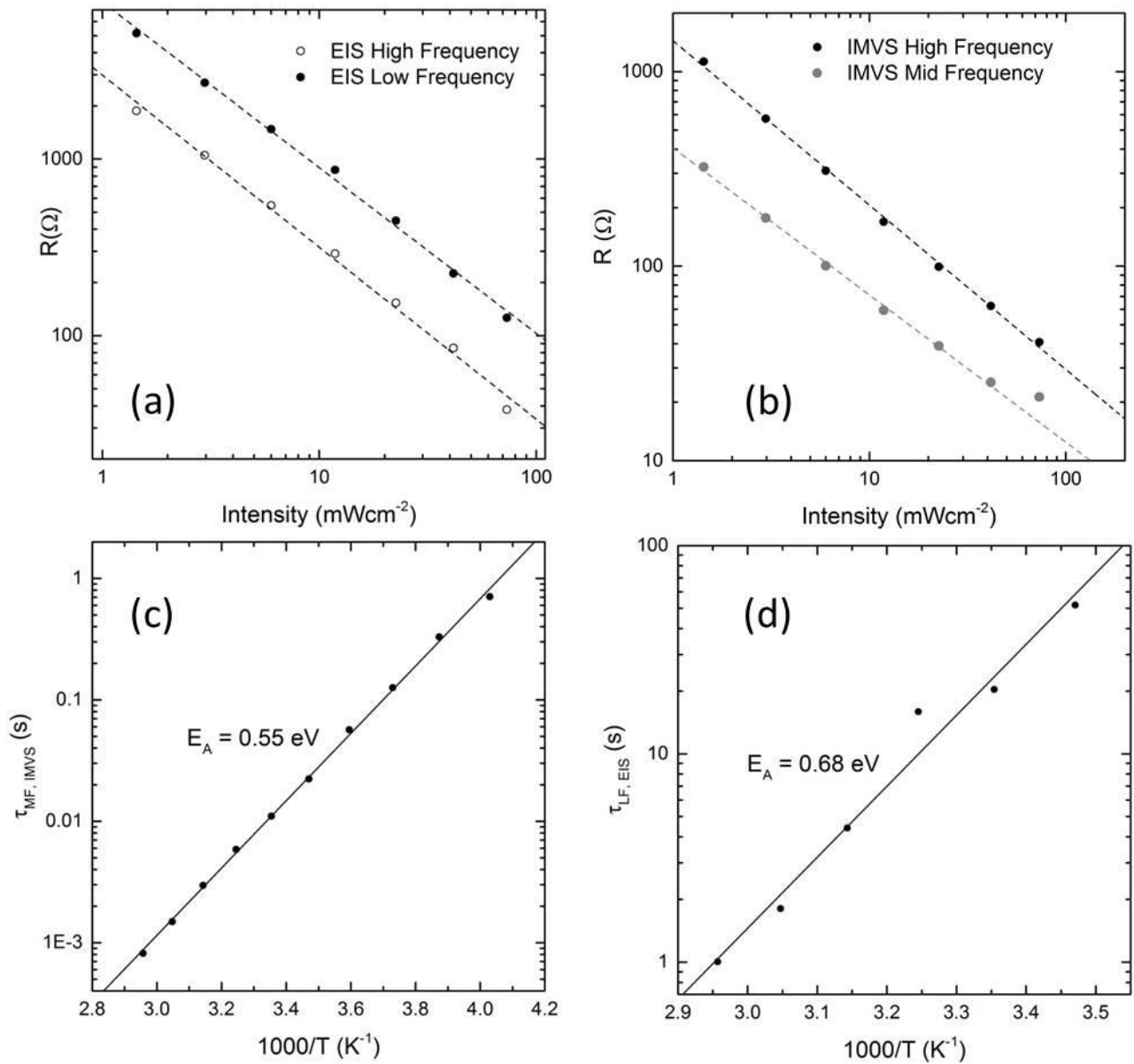

Figure 6. (a) Intensity dependence of the resistances extracted from the high frequency and low frequency semi-circles measured by EIS on cell CT-3 (b) Intensity dependence of the resistance values extracted from the HF and MF semi-circles observed in IMVS (scaled to give the same value of Cgeo, cell EY-4). (c) Temperature dependence of the mid frequency time constant measured from IMVS. (d) Temperature dependence of the low frequency time constant measured by EIS.

The low frequency feature with an associated activation enthalpy of 0.68 eV and a lower attempt frequency of $10^{10}$ Hz could be associated with the motion of $MA^+$ vacancies. The low attempt frequency would suggest that there is significant negative entropy of activation associated with the movement of $MA^+$ vacancies. This would be the case if $MA^+$ can only move from cage to cage when it is in certain preferred orientations. A value of 0.68 eV is lower than the value of 0.84 eV predicted by Eames et al. However Meloni et al,[9] Azpiroz et al[7] and Harayama et al[8] have all predicted slightly lower activation energies for $MA^+$ vacancy migration. It cannot be ruled out, however, that the two activation energies are for vacancy movement in two distinct environments, e.g. movement of iodide vacancies in the bulk compared to movement along grain boundaries.

As outlined above, most EIS studies of perovskite solar cells (planar and mesostructured) report a low frequency relaxation process. The most prevalent explanations for the low frequency feature are a giant dielectric effect,[27] ionic diffusion in the bulk [32] or charge accumulation at an interface. [10,21] Bag et al have suggested that the low frequency response is due to a pure ionic impedance.[32] They measured planar perovskite devices with organic (PEDOT:PSS and PCBM) contacts. The low frequency response was modelled with a shorted Warburg element to describe diffusion of ions through the bulk perovskite. A shorted Warburg model only applies if the ions can move through the perovskite-ETM and perovskite-HTM interfaces. The low frequency intercept with the real axis in the Nyquist plot then gives the value of the 'ion transfer resistance'. In their case the measurements were terminated at 100 Hz, but the ion transfer resistance would have had a value of several thousand Ohm at 100 mWcm$^{-2}$. The physical origin of the shorted Warburg is not explained in the paper. The temperature dependence of the low frequency time constant was used to extract an activation energy of 0.58 eV between 320 K and 300K which was attributed to $MA^+$ vacancy migration. Values of the shorting resistance and the intensity dependence of the data were not reported.

Other authors have suggested that the low frequency response is electronic in nature.[14,27,41] In this approach the low frequency semi-circle is modelled using a resistor and capacitor in parallel. In early papers the low frequency feature was attributed to a giant dielectric effect where the dielectric constant at 0.1 Hz was found to be $10^3$ in the dark and $10^6$ under illumination.[27] The increase was attributed to alignment of the $MA^+$ cation with the field. An alternative interpretation is to assume that a very large capacitance ( ≥ mF cm$^{-2}$) is measured at low frequency. It has been argued that this giant capacitance is too large to be due to an ionic double layer and is instead due to a charge accumulation layer. The approach of Zarazua et al links the slow formation of ionic defects with charge accumulation at the contact.[31] This model suggests a giant capacitance that increases rapidly with the photogeneration of carriers. The associated values of the resistance have not been studied in detail; indeed many publications plot only the complex capacitance as a function of frequency and ignore the value of the resistance obtained from the low frequency intercept.[44] This model also fails to explain why the HF recombination resistance and the LF (and MF) resistances would show near identical light intensity dependences. In addition our temperature dependent data gives activation energies that seem reasonable for vacancy migration/diffusion. Defect formation, on the other hand, has been calculated to occur extremely easily with calculated activation energies as low as 0.14 eV.[3]

All of the existing models fail to fully explain the impedance response of PSC. Here we present a new explanation that describes all of the features seen in EIS and IMVS and is also consistent with the OCVD measurements outlined above. We suggest that the MF and LF relaxation processes are due to a coupled electronic-ionic impedance e.g. where the distribution of ions in the perovskite is able to modify the rate of charge carrier recombination. This interdependence between ionic distribution and charge extraction/recombination has been predicted by Van Reenen et al[4], Frost and Walsh[3] as well as investigated experimentally by Calado et al[16]. In EIS measurements the recombination resistance is therefore a *complex frequency dependent parameter* due to the motion of vacancies on the ms and s timescales. Two low frequency semi-circles are seen because the recombination resistance increases as the ionic vacancies relax - first on the ms time scale (to give the MF feature) and then on the s time scale (to give the LF feature). Small changes in recombination resistance are enough to explain the effect and it was seen above that quite large changes in recombination are possible as the ionic environment relaxes. At low frequencies (long relaxation times) the total recombination resistance only increases from ~ 50 Ω to ~ 200 Ω at the highest light intensity.

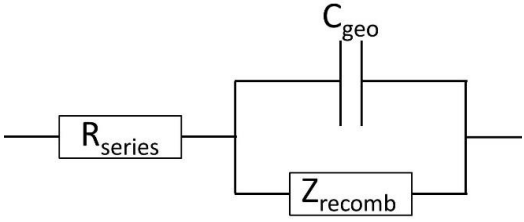

Figure 7. Simplified equivalent circuit model for PSC which is used to explain the frequency dependent response. The model contains the series resistance $R_{series}$, the geometric capacitance $C_{geo}$ and the frequency dependent recombination resistance $Z_{recomb}$.

The simplified model outlined in Figure 7 is entirely consistent with the full model shown in Figure 3. When resistors are in parallel the smallest resistance will always dominate the response, so the shunt resistance is typically only measured at extremely low light intensities where $Z_{recomb} > R_{shunt}$. In the case of the cells measured here at the lowest light intensity (1.5 mW cm$^{-2}$) $R_{recomb}$ was only ~2000 Ω which remains considerably below $R_{shunt}$ ~ 4 x 10$^5$ Ω (measured in ref 17). The time constant $R_{ionic}C_{dl}$ would likewise only appear in the EIS/IMVS plots when the magnitude of $Z_{recomb}$ becomes comparable to $R_{ionic}$. As outlined above, if $\tau_{OCVD,2} = R_{ionic}C_{dl}$, then taking a time constant of 5 s (at 25°C, 73 mWcm$^{-2}$) and a calculated double layer capacitance of 40 μF cm$^{-2}$ it would give an ionic resistance of 1.25 x 10$^5$ Ω cm$^2$. This is substantially above the values of $R_{recomb}$ measured at all light intensities in this study and simple simulations using the equivalent circuit outlined in Figure 3 and experimentally relevant values of the resistances and capacitances can be used to demonstrate that $R_{ionic}C_{dl}$ would not be measured by EIS.

The data underlines the fact that at long times OCVD is measuring a different process to EIS/IMVS. OCVD is probing charging of the ionic double layer after the light is switched off. EIS and IMVS are measuring small frequency dependent changes in the recombination resistance as the distribution of ionic vacancies is perturbed by the modulated field/light.

Our simple model explains the remarkably similar intensity dependence of the HF, MF and LF resistances, as well as the fact that the temperature dependence of the time constants for the MF and LF processes give very reasonable activation energies for thermally activated motion of ionic vacancies. To explain in more detail: the potential drop across the ionic double layers in the PSC will be dependent on the concentration and distribution of the vacancies close to the contacts. In EIS and IMVS measurements, the voltage difference across the cell is modulated (either externally or internally), which will lead to a corresponding perturbation of vacancies in the double layers and modification of the recombination resistance. The changes in recombination resistance will occur on the time scale of ion movement e.g. at low frequencies when the ions interact with the oscillating field.

At open circuit, the relaxation of vacancies is beneficial for the PSC since the recombination resistance actually *increases*. This could occur if relaxation of vacancies leads to better screening of electronic charge in the perovskite layer or more favourable band offsets for extraction. The increase in recombination resistance with frequency is quite small – a factor of 4-5 at the highest light intensity. It is reasonable that slight adjustments to the barrier height for recombination or to the degree of ion-electron screening would be enough to explain the effect.

It is important to underline that although the MF and LF time constants are dictated by ion movement, a pure ionic impedance is not being measured; e.g. at open circuit in these devices we *are not* measuring a Warburg impedance due to ion diffusion. It is also important to note that if we are measuring a complex recombination resistance then it is not reasonable to extract values of capacitance from the MF and LF relaxation processes. A semi-circular response is seen simply because the recombination resistance changes with frequency, but the associated capacitance does not have physical meaning.

There are several possible explanations for the observation of two lower frequency features (MF and LF) in our measurements. Firstly the recombination resistance could be modified by interaction with two different ionic species as we have outlined already above. If we assume that relaxation of the ionic charge distribution occurs predominantly by diffusion as a consequence of the steep gradients in concentration near the contacts, then it is possible to estimate the diffusion length of the ions from the simple relationship

$$L_d = \sqrt{D\tau} \qquad (3)$$

Taking diffusion coefficients of 10$^{-16}$ cm$^2$s$^{-1}$ and 10$^{-12}$ cm$^2$s$^{-1}$ and $\tau_{MF}$ (IMVS) of 15 ms at 73 mWcm$^{-2}$;[6] it would suggest diffusion distances of 1.2 nm for I$^-$ (or 12 pm for

MA$^+$). A lifetime of $\tau_{LF}$ from IMVS of 12s would translate to a diffusion length of 0.35 nm for MA$^+$ vacancies (or 35 nm for I$^-$). It seems very possible that the diffusion length for the I$^-$ and MA$^+$ vacancies in the double layer are 1.2 nm and 0.35 nm respectively in response to the small amplitude perturbation. There are other possible reasons why two time constants have been measured. There could be two types of I$^-$ (or MA$^+$) vacancies with different diffusion rates, for example defects in the bulk of the perovskite and defects around the crystallite surface. We could be looking at the movement of vacancies compared to interstitial defects. Alternatively the recombination resistance could be modified by faster ion migration under a field and slower ion diffusion in the bulk where there is no field.

**Conclusions**

We have measured temperature and intensity dependent EIS, IMVS and photovoltage transients for PSC. Our results strongly suggest that the recombination resistance in PSC depends on the ionic environment. Ionic vacancies move to increase the recombination resistance on the ms to s timescale, which we suggest is due to a reduction in capture cross section as electronic charge is screened and/or ionic modulation of the band offsets in the device. It is also possible that ionic vacancies can stabilize filled traps, hence effectively reducing the vacant trap density which slows down the non-radiative recombination rate. We propose a simple model which can explain cell behaviour in both the frequency and the time domain; key to the model is this fact that the recombination resistance is frequency dependent. The low frequency response in EIS and IMVS measurements appears as a semi-circle in the Nyquist plot simply because the recombination resistance changes as ionic vacancies relax.

Our conclusion is supported by the fact that the temperature dependence of the MF and LF lifetimes from EIS and IMVS gives activation energies of 0.55 eV and 0.68 eV which are consistent with computationally predicted values for thermally activated ion movement. The activation energy for ion movement can also be obtained from the photovolatge rise and OCVD measurements to give values of 0.42 and 0.58 eV; again suggesting that the photovoltage rise and OCVD are being strongly influenced by the ionic environment. At short times the OCVD response is dominated by discharge of the geometric capacitance across the recombination resistance. On the ms time scale a bounce back in the OCVD response is seen as changes in the ionic environment affect the recombination resistance. The extremely slow decay of the remaining photovoltage can be explained if it is caused by charging of the ionic double layer.

As discussed in the introduction it is increasingly clear that both mobile ions and the nature of ETL and HTL used as contacts are responsible for jV hysteresis and photovoltage rise times in PSC. This observation is consistent with our results. Ionic modification of band offsets and local band bending will depend on the nature of the contact. To give one example - the locus of specifically adsorbed ions at the perovskite-ETL and perovskite-HTL interfaces will depend on whether ions are confined to the perovskite (in the case of 'hard' inorganic contacts) or whether they can pass into the contact (e.g. in the case of a soft polymer contact). Different contacts could therefore substantially change the interfacial ion distribution and the exact location of the double layer which would in turn strongly influence the recombination rate. The voltage rise time under illumination, as well as the EIS response would therefore be highly contact dependent. Significantly we show that EIS provides a way of quantifying how much small changes in the distribution of ionic vacancies can modify the recombination resistance. The techniques allows a direct comparison of devices prepared with different contacts and using different preparation methods.

**Acknowledgements**

PJC would like to thank Aron Walsh, Jarvist Frost and Saiful Islam for useful discussions. PJC and AP acknowledge funding from the EPSRC Doctoral Training Centre in Sustainable Chemical Technologies: EP/G03768X/1

**Notes and references**

1. http://www.nrel.gov/ncpv/images/efficiency_chart.jpg.

2. C. C. Stoumpos and M. G. Kanatzidis, *Adv Mater*, 2016, **28**, 5778-5793.

3. J. M. Frost and A. Walsh, *Acc. Chem. Res.*, 2016, **49**, 528-535.

4. S. van Reenen, M. Kemerink and H. J. Snaith, *J. Phys. Chem. Lett.*, 2015, **6**, 3808-3814.

5. G. Richardson, S. E. J. O'Kane, R. G. Niemann, T. A. Peltola, J. M. Foster, P. J. Cameron and A. B. Walker, *Energy Environ. Sci.*, 2016, **9**, 1476-1485.

6. C. Eames, J. M. Frost, P. R. F. Barnes, B. C. O/'Regan, A. Walsh and M. S. Islam, *Nat Commun*, 2015, **6**.

7. J. M. Azpiroz, E. Mosconi, J. Bisquert and F. De Angelis, *Energy Environ. Sci.*, 2015, **8**, 2118-2127.

8. J. Haruyama, K. Sodeyama, L. Y. Han and Y. Tateyama, *J. Am. Chem. Soc.*, 2015, **137**, 10048-10051.

9. S. Meloni, T. Moehl, W. Tress, M. Franckevicius, M. Saliba, Y. H. Lee, P. Gao, M. K. Nazeeruddin, S. M. Zakeeruddin, U. Rothlisberger and M. Graetzel, *Nat Commun*, 2016, **7**.

10. H. S. Kim, I. H. Jang, N. Ahn, M. Choi, A. Guerrero, J. Bisquert and N. G. Park, *J. Phys. Chem. Lett.*, 2015, **6**, 4633-4639.

11. A. Guerrero, G. Garcia-Belmonte, I. Mora-Sero, J. Bisquert, Y. S. Kang, T. J. Jacobsson, J.-P. Correa-Baena and A. Hagfeldt, *J. Phys. Chem. C*, 2016, **120**, 8023-8032.

12. Y. Han, S. Meyer, Y. Dkhissi, K. Weber, J. M. Pringle, U. Bach, L. Spiccia and Y.-B. Cheng, *J. Mater. Chem. A*, 2015, **3**, 8139-8147.

13. Y. Kato, L. K. Ono, M. V. Lee, S. Wang, S. R. Raga and Y. Qi, *Adv. Mater. Int.*, 2015, **2**, n/a-n/a.

14. A. Guerrero, J. You, C. Aranda, Y. S. Kang, G. Garcia-Belmonte, H. Zhou, J. Bisquert and Y. Yang, *ACS Nano*, 2016, **10**, 218-224.

15. T.-Y. Yang, G. Gregori, N. Pellet, M. Grätzel and J. Maier, *Angew. Chem.*, 2015, **127**, 8016-8021.

16. P. T. Calado, Andrew M.; Bryant, Daniel; Li, Xiaoe; Nelson, Jenny; O'Regan, Brian C.; Barnes, Piers R. F., 2016.

17. S. Casaluci, L. Cinà, A. Pockett, P. S. Kubiak, R. G. Niemann, A. Reale, A. Di Carlo and P. J. Cameron, *J. Power Sources*, 2015, **297**, 504-510.

18. A. Pockett, G. E. Eperon, T. Peltola, H. J. Snaith, A. Walker, L. M. Peter and P. J. Cameron, *J. Phys. Chem. C*, 2015, **119**, 3456-3465.

19. N. Sakai, S. Pathak, H.-W. Chen, A. A. Haghighirad, S. D. Stranks, T. Miyasaka and H. J. Snaith, *J. Mater. Chem. A*, 2016, **4**, 4464-4471.

20. L. Bertoluzzi, R. S. Sanchez, L. F. Liu, J. W. Lee, E. Mas-Marza, H. W. Han, N. G. Park, I. Mora-Sero and J. Bisquert, *Energy Environ. Sci.*, 2015, **8**, 910-915.

21. N. W. Duffy, L. M. Peter, R. M. G. Rajapakse and K. G. U. Wijayantha, *Electrochem. Commun.*, 2000, **2**, 658-662.

22. J. Kruger, R. Plass, M. Gratzel, P. J. Cameron and L. M. Peter, *J. Phys. Chem. B*, 2003, **107**, 7536-7539.

23. S. D. Stranks, V. M. Burlakov, T. Leijtens, J. M. Ball, A. Goriely and H. J. Snaith, *Phys Rev Appl*, 2014, **2**.

24. Y. Yamada, T. Nakamura, M. Endo, A. Wakamiya and Y. Kanemitsu, *J. Am. Chem. Soc.*, 2014, **136**, 11610-11613.

25. A. Baumann, K. Tvingstedt, M. C. Heiber, S. Väth, C. Momblona, H. J. Bolink and V. Dyakonov, *APL Materials*, 2014, **2**, -.

26. R. S. Sanchez, V. Gonzalez-Pedro, J.-W. Lee, N.-G. Park, Y. S. Kang, I. Mora-Sero and J. Bisquert, *J. Phys. Chem. Lett.*, 2014, **5**, 2357-2363.

27. E. J. Juárez-Pérez, R. S. Sánchez, L. Badia, G. Garcia-Belmonte, Y. S. Kang, I. Mora-Sero and J. Bisquert, *J. Phys. Chem. Lett.*, 2014, **9**, 2390-2394.

28. V. Gonzalez-Pedro, E. J. Juarez-Perez, W.-S. Arsyad, E. M. Barea, F. Fabregat-Santiago, I. Mora-Sero and J. Bisquert, *Nano Lett.*, 2014, **14**, 888-893.

29. J. Bisquert, L. Bertoluzzi, I. Mora-Sero and G. Garcia-Belmonte, *J. Phys. Chem. C*, 2014, **118**, 18983-18991.

30. E. Guillén, F. J. Ramos, J. A. Anta and S. Ahmad, *The Journal of Physical Chemistry C*, 2014.

31. I. Zarazua, J. Bisquert and G. Garcia-Belmonte, *J. Phys. Chem. Lett.*, 2016, **7**, 525-528.

32. M. Bag, L. A. Renna, R. Y. Adhikari, S. Karak, F. Liu, P. M. Lahti, T. P. Russell, M. T. Tuominen and D. Venkataraman, *J. Am. Chem. Soc.*, 2015, **137**, 13130-13137.

33. A. R. Pascoe, M. J. Yang, N. Kopidakis, K. Zhu, M. O. Reese, G. Rumbles, M. Fekete, N. W. Duffy and Y. B. Cheng, *Nano Energy*, 2016, **22**, 439-452.

34. Q. D. Tai, P. You, H. Q. Sang, Z. K. Liu, C. L. Hu, H. L. W. Chan and F. Yan, *Nat. Comm.*, 2016, **7**.

35. K. Miyano, N. Tripathi, M. Yanagida and Y. Shirai, *Acc. Chem. Res.*, 2016, **49**, 303-310.

36. P. J. Cameron and L. M. Peter, *J. Phys. Chem. B*, 2003, **107**, 14394-14400.

37. G. Schlichthoerl and L. M. Peter, *J. Electrochem. Soc.*, 1995, **142**, 2665-2669.

38. E. A. Ponomarev and L. M. Peter, *J. Electroanal. Chem.*, 1995, **397**, 45-52.

39. A. Dualeh, T. Moehl, N. Tétreault, J. Teuscher, P. Gao, M. K. Nazeeruddin and M. Grätzel, *ACS Nano*, 2014, **8**, 362-373.

40. A. R. Pascoe, N. W. Duffy, A. D. Scully, F. Z. Huang and Y. B. Cheng, *J. Phys. Chem. C*, 2015, **119**, 4444-4453.

41. O. Almora, I. Zarazua, E. Mas-Marza, I. Mora-Sero, J. Bisquert and G. Garcia-Belmonte, *J. Phys. Chem. Lett.*, 2015, **6**, 1645-1652.

42. A. R. Pascoe, M. Yang, N. Kopidakis, K. Zhu, M. O. Reese, G. Rumbles, M. Fekete, N. W. Duffy and Y.-B. Cheng, *Nano Energy*, 2016, **22**, 439-452.

43. H.-S. Kim, I. Mora-Sero, V. Gonzalez-Pedro, F. Fabregat-Santiago, E. J. Juarez-Perez, N.-G. Park and J. Bisquert, *Nat. Comm.*, 2013, **4**, 2242-2242.

44. H.-S. Kim, I.-H. Jang, N. Ahn, M. Choi, A. Guerrero, J. Bisquert and N.-G. Park, *J. Phys. Chem. Lett.*, 2015, **6**, 4633-4639.